\documentclass[prfluids]{revtex4-2}
\usepackage[utf8]{inputenc}
\usepackage[T1]{fontenc}
\usepackage{graphicx}% Include figure files
\usepackage{dcolumn}% Align table columns on decimal point
\usepackage{bm,bbold}% bold math
\usepackage{amsmath,amssymb}
\usepackage{dsfont}
\usepackage{physics}
\usepackage{siunitx}

\usepackage{booktabs}
\usepackage{adjustbox}
\usepackage{xcolor}
\usepackage{hyperref}
\hypersetup{ 
    colorlinks=true,
    citecolor=blue,
    linkcolor=violet,
    urlcolor=magenta,
    bookmarksopen=true,
    linktoc=page
 }

\newcommand{\code}[1]{\texttt{\textbf{\textcolor{black}{#1}}}}

\usepackage[capitalise]{cleveref}
\crefrangeformat{equation}{Eqs.~(#3#1#4-#5#2#6)}
\crefformat{section}{Sec.~#2#1#3}
\crefmultiformat{section}{Secs.~#2#1#3}{ and~#2#1#3}{, #2#1#3}{ and~#2#1#3}

\newcommand{\eps}{\varepsilon}
\newcommand{\mum}{\unit{\micro\meter}}

\begin{document}

\title{
    Polydisperse collision kernels in droplet-laden turbulence with implications for rain formation
}

\author{Lukas A.
    Codispoti} \email{lukasco@ethz.ch} \author{Daniel W.
    Meyer} \author{Patrick Jenny} \affiliation{ Institute of Fluid Dynamics, ETH
    Z\"urich\\Sonneggstrasse 3, 8092 Z\"urich, Switzerland} \date{\today}

\begin{abstract}
    The collision kernel of droplets in warm
    clouds is a crucially important quantity for the parameterization of
    precipitation in weather and climate models.
    Nevertheless, its accurate representation remains a challenge, specifically in
    the bottleneck range $15\,\mum<r<40\,\mum$, within which turbulence is believed to substantially contribute to droplet growth.
    In this work, we address this problem by performing direct numerical
    simulations of polydisperse inertial particles suspended in three-dimensional
    turbulence at Reynolds number up to $Re_\lambda=418$.
    Collision statistics are compiled for droplet pairs across the Stokes number
    range $St\in[0.02,2]$, yielding comprehensive bidisperse maps of collision
    kernels, radial relative velocities, and radial distribution functions at
    contact.
    Our analysis reveals that polydispersity enhances collisions between light
    droplets through differential sampling, but attenuates collisions at larger
    Stokes numbers by rapidly reducing the spatial overlap of droplet clusters.
    By benchmarking existing models, we show that dominant bidisperse errors
    arise from overpredicted cross-species clustering.
    In light of these results, we propose an adapted model for the bidisperse
    radial distribution function, as well as a
    novel parameterization for the associated collision kernel,
    applicable to the smallest droplets in the bottleneck range with small
    settling velocities.
    % We compare our results for bidisperse clustering and the collision kernel to
    % the existing models of Zhou et al.~[\href{https://doi.org/10.1017/S0022112000003372}{J.~Fluid~Mech.~\textbf{433},~77~(2001)}] and Onishi~\&~Seifert
    % [\href{https://doi.org/10.5194/acp-16-12441-2016}{Atmos.~Chem.~Phys.~\textbf{16},~12441~(2016)}], respectively.
    % In light of these comparisons, an adapted model for the bidisperse radial
    % distribution function is proposed, as well as a
    % novel parameterization for the associated collision kernel,
    % applicable to the smallest droplets in the bottleneck range with small
    % settling velocities.
    Finally, we study the broadening of the droplet size distribution due to
    collision-coalescence and demonstrate that droplet growth 
    is markedly accelerated in parcels of large local dissipation rate, 
    supporting the hypothesis that turbulent intermittency may help overcome
    the bottleneck barrier.
\end{abstract}

\flushbottom
\maketitle
\thispagestyle{empty}

\section{Introduction}\label{sec:introduction}
Cumulus clouds form in the shallow regions of Earth's atmosphere when moist air
ascends and cools, causing water vapor to condense at nucleation sites into
micron-sized droplets that subsequently grow and, eventually, fall to the
ground as rain.
Despite this seemingly simple picture, key aspects of cloud-droplet growth
remain poorly understood, contributing to persistent uncertainty in weather and
climate predictions~\cite{bodenschatzCanWeUnderstand2010}.
One long-standing question is how droplets with radii
$15\,\mum < r < 40\,\mum$ in warm (ice-free) clouds can grow rapidly enough to
initiate precipitation
within roughly 30 minutes, since neither
condensational growth nor collision-coalescence driven by gravitational
settling is sufficiently effective in this so-called \emph{bottleneck} range~\cite{grabowskiGrowthCloudDroplets2013}.
Increasing evidence suggests that turbulence, the chaotic motion of fluid
flow, plays an important role in resolving this puzzle.
The broad range of turbulent flow scales alters the spatial distribution of
droplets and amplifies relative velocities, thus increasing
collision-coalescence rates compared to quiescent conditions.
These enhanced rates are quantified by collision kernels, which describe how
frequently droplets of given sizes encounter and collide.
Accurate models for the collision kernel are essential for cloud microphysics
schemes in weather and climate simulations, as they directly govern the
evolution of the droplet size distribution (DSD).
Developing reliable parameterizations nevertheless remains challenging due to
the limited theoretical understanding of droplet-turbulence interactions,
scarce in situ observations, and numerical constraints.

% Cloud droplets suspended in turbulent air behave as small inertial particles,
% whose finite response time,
Warm clouds are dilute suspensions of small droplets, whose
motion can be approximated by that of inertial particles in turbulence.
Their finite response time
\begin{equation}\label{eqn:taup}
  \tau_p=\frac{2r^2\rho_p}{9\nu\rho_f},
\end{equation}
leads to dynamics that differ from those of passive fluid tracers.
Here, $\nu$ is the kinematic viscosity of the flow, and $\rho_p$ and $\rho_f$
denote the density of the particle and the flow, respectively.
There exists a large body of literature on the motion of inertial particles in
turbulence (e.g.,
Refs.~\cite{maxeyGravitationalSettlingAerosol1987,squiresPreferentialConcentrationParticles1991,cenciniDynamicsStatisticsHeavy2006,rayPreferentialConcentrationRelative2011}).
In the context of droplet collisions, turbulence acts primarily through three
closely related mechanisms: preferential concentration, which alters the spatial
distribution of droplets; polydispersity, which enhances relative motion between
droplets with different $\tau_p$ through differential sampling of the flow;
and the sling effect, whereby droplet trajectories cross at large relative
velocities in rare extreme events.

Preferential concentration arises due to the interaction of inertial particles
with turbulent vortices, leading to non-uniform spatial distributions, also
referred to as \emph{clustering}.
Particles tend to be expelled from regions of large vorticity and accumulate in
strain-dominated areas, creating a compressible effective flow of particles~\cite{maxeyGravitationalSettlingAerosol1987}.
The inertia of a particle is quantified by the Stokes number
$St=\tau_p/\tau_\eta$, where
$\tau_\eta = \sqrt{\nu/\eps}$ is the Kolmogorov time, characterizing the
timescale of the dissipative scales, and $\eps$ is the turbulent kinetic energy
dissipation rate.
A particle with $St\ll1$ behaves like a fluid tracer, whereas with $St\gg1$,
particles move ballistically and decorrelate from the carrier flow.
In both limits, particles disperse uniformly in space.
In between the two extremes and most notably around $St\approx1$, on the other
hand, particles cluster (e.g.,
Refs.~\cite{squiresPreferentialConcentrationParticles1991,cenciniDynamicsStatisticsHeavy2006,rayPreferentialConcentrationRelative2011}), which naturally increases collision
rates between inertial by increasing the likelihood of encounters in regions of
high number density.

Polydispersity, the presence of particles with varying sizes, introduces an
additional mechanism affecting collisions.
Because particles of different sizes interact differently with turbulent
structures, their trajectories decorrelate, leading to increased relative
velocities.
At the same time, differently sized particles tend to cluster in different
regions of the domain, thereby reducing inter-species preferential
concentration.
From a phenomenological standpoint, polydispersity merits investigation since
cloud droplets are inherently size-distributed in nature.
While this mechanism has been studied numerically in two-dimensional turbulence
(e.g., Ref.~\cite{jamesEnhancedDropletCollision2017}) and three-dimensional
turbulence at low Reynolds numbers (e.g.,
Refs.~\cite{nairEffectGravityParticle2023,zhouModellingTurbulentCollision2001}),
comprehensive three-dimensional analyses at high Reynolds numbers remain
limited due to computational constraints.
In the present work, we extend these efforts by conducting a detailed
investigation of polydisperse collision dynamics in three-dimensional
turbulence at high Reynolds number.

Finally, the sling effect~\cite{falkovichAccelerationRainInitiation2002} is the
mechanism through which particles experience rare extreme turbulent events and
are ``slung'' out of vortices.
This phenomenon occurs when particles from different flow regions cross
trajectories, creating singularities in phase space, so-called \emph{caustics}~\cite{wilkinsonCausticsTurbulentAerosols2005}.
Caustic formation naturally leads to collisions between particles that travel
on different branches of the phase-space manifold, i.e., exhibit different
velocities while occupying the same location in physical space.
The occurrence of caustics implies that inertial particles do not move on a
continuum; thus, there is no well-defined Eulerian field that can describe the
motion of the particles.
Caustics in turbulent suspensions have been confirmed both
numerically~\cite{meibohmCausticFormationNonGaussian2024,codispotiDissectingInertialClustering2025}
and experimentally~\cite{bewleyObservationSlingEffect2013}.
Yet, whether the sling effect is effective for the collision-coalescence of
cloud droplets has been the subject of debate in recent years~\cite{deepuCausticsinducedCoalescenceSmall2017,fouxonIntermittencyCollisionsFast2022,ravichandranWaltzTinyDroplets2022}.

A central question is how the effects of turbulence and gravitational settling
compete in inducing droplet collisions.
Their relative importance is given by the settling parameter
$Sv=\tau_p g /u_\eta$, where $g$ is the gravitational acceleration
and $u_\eta=(\eps\nu)^{1/4}$ is the Kolmogorov velocity.
% , $u_\eta=\eta/\tau_\eta$ is the Kolmogorov velocity scale, and $\eta$ denotes
% the Kolmogorov length.
Under typical conditions in cumulus clouds, the mean dissipation rate is moderate,
$\langle \eps \rangle \le 0.05\,\unit{m^2/s^{3}}$~\cite{siebertHighresolutionMeasurementCloud2015}.
For droplets with radii $r\gtrsim15\,\mum$, this corresponds to $Sv>1$, so that
gravitational settling is more important relative to turbulence.
Importantly, however, this does not imply rapid collisional growth: in the
bottleneck range, both turbulence-induced inertial effects and gravitational
settling remain weak in absolute terms.

This picture changes when one moves away from mean conditions and considers the
strong spatial variability in clouds.
Measurements in atmospheric clouds are becoming available at
ever-increasing resolutions, revealing evidence that a fraction of
droplets experiences local conditions conducive to inertial
clustering and caustic formation, even when average conditions would suggest
otherwise~\cite{siebertHighresolutionMeasurementCloud2015,thiedeHighlyLocalisedDroplet2025}.
One natural source of such conditions is intermittency in the local energy
dissipation rate.
Since $St \varpropto r^2\eps^{1/2}$ and $Sv\varpropto r^2\eps^{-1/4}$,
inertial effects increase with $\eps$, while reducing the relative importance
of settling.
Physically, the settling velocity of a droplet is unchanged, whereas the
characteristic turbulent velocity increases; as a result, turbulence can become
dynamically important in sufficiently intense local turbulence even for droplets
that, under mean conditions, would appear to be governed primarily by gravity.
Observations support the prevalence of such heterogeneous conditions. 
For example, Thiede et al.~\cite{thiedeHighlyLocalisedDroplet2025} report
pronounced inertial clustering within parcels on the order of
centimeters, while Allwayin et al.~\cite{allwayinLocallyNarrowDroplet2024}
reveal that local DSDs can be unexpectedly narrow, deviating significantly from integral-scale averages.
Accelerated growth of a small subset of fortuitous droplets experiencing
rare intense turbulence may also connect to the
\emph{lucky droplet hypothesis}~\cite{kostinskiFluctuationsLuckDroplet2005},
according to which warm-rain
initiation is controlled not by mean statistics but by the fat tails of the
relevant probability distributions.
Understanding collision dynamics in this intermittent regime is therefore
highly relevant to the bottleneck problem.

In this study, we perform an extensive DNS campaign of fully-resolved,
three-dimensional particle-laden turbulence at $Re_\lambda=55$, $179$ and $418$.
Our aim is to elucidate the turbulent mechanisms governing droplet collisions in
the bottleneck regime of warm-cloud microphysics.
In particular, we provide what is, to our knowledge, the most extensive map of
the bidisperse collision kernel obtained from high-Reynolds-number DNS to date,
covering droplet pairs within the range $St\in[0.02,2]$.
We compare our results to existing models and propose modifications for the
expression of bidisperse clustering as well as a novel parameterization for the
bidisperse collision kernel.
To isolate the turbulent enhancement of droplet collisions, gravity is neglected
throughout much of this work.
In~\Cref{sec:simulations_with_gravity}, however, it is reintroduced,
and we show that the proposed parameterization remains
qualitatively predictive for settling droplets
within the bottleneck range.
Finally, we examine the evolution of the DSD and the impact of locally enhanced
dissipation in simulations of coalescing, settling droplets.

Before we begin, a few remarks are warranted regarding the influence of the
Reynolds number and turbulent intermittency.
Atmospheric clouds exhibit Taylor-scale Reynolds numbers of
$\mathcal{O}(Re_\lambda)=10^4$, which are unattainable in DNSs with today's
computing capabilities (fully resolving turbulence at
$Re_\lambda=10^4$ would require $\mathcal{O}(10^2\,\unit{PB})$ of memory).
A systematic variational analysis of the impact of the Reynolds number is beyond
the scope of this work.
Prior work shows $Re_\lambda$-insensitivity of clustering for $St\le0.2$,
modest attenuation for $0.4\le St\le 1$, and enhancement for $St\ge 2$ as
turbulent intermittency increases with the Reynolds
number~\cite{onishiCollisionStatisticsInertial2014,onishiLagrangianTrackingSimulation2015,onishiReynoldsnumberDependenceTurbulence2016}.
Similar trends appear in the collision
statistics~\cite{irelandEffectReynoldsNumber2016}.
The results suggest that $Re_\lambda$ mainly affects cloud microphysics by
reshaping the distribution of the local dissipation (its intermittency), not by
altering the small-scale mechanics at fixed $St$.
In other words, the dominant influence of the Reynolds number is statistical
through the intermittency of $\eps$ rather than dynamical at the droplet-scale.
In this work, we compute the collision statistics within simulation domains
that correspond to physical cloud parcels on the scale of decimeters.
At fixed volume-averaged dissipation rate, the collision
dynamics within these control volumes are largely insensitive to $Re_\lambda$.

\section{Turbulent collision kernel}
\label{sec:turbulent_collision_kernel}
We consider collisions between droplet pairs $(i,j)$ with radii $(r_i,r_j)$,
separated by the distance $\bm{\Delta x} = \bm{x}_i-\bm{x}_j$ with the relative
velocity $\bm{\Delta v}=\bm{v}_i-\bm{v}_j$.
The number of collisions per unit time is given by the collision rate
$\mathbb{N}_{ij}$, and the dynamic collision kernel
\begin{equation}
    \label{eqn:CK} \Gamma_{ij} = \frac{\mathbb{N}_{ij}}{n_in_j}
\end{equation}
is
thence obtained by normalizing with the respective number densities, $n_i$ and
$n_j$, 
with an additional factor of $2$ in the monodisperse case ($i=j$).
The widely used kinematic formulation~\cite{sundaramCollisionStatisticsIsotropic1997a,wangStatisticalMechanicalDescriptions1998}
\begin{equation}
    \label{eqn:SC} \Gamma = 2\pi R^2\langle w_r \rangle g(R)
\end{equation}
serves as the basis for most modeling approaches.
Here,
\begin{equation}
    \label{eq:wr} w_r = \frac{|\bm{\Delta v} \cdot
        \bm{\Delta x}|}{|\bm{\Delta x}|} \bigg|_{|\bm{\Delta x}|=R}
\end{equation}
is
the radial relative velocity (RRV) and $g(R)$ is the radial distribution
function (RDF) at contact distance $R=r_i+r_j$.
The RDF quantifies the degree of preferential concentration.
Its bidisperse form is symmetric ($g_{ij}(r)=g_{ji}(r)$) and is computed as
\begin{equation}
    \label{eqn:rdf} g_{ij}(r) = \frac{\langle N_j\rangle_i(r) /
        V_{s}}{N_{ij} V},
\end{equation}
where $\langle N_j\rangle_i(r)$ is
the mean number of particles of species $j$ at separation $r$ from a particle of
species $i$ within the spherical shell of volume $V_{s} =
    \tfrac{4\pi}{3}[(r+s/2)^3-(r-s/2)^3]$ and thickness $s$, $N_{ij}$ is
the total number of $(i,j)$ pairs, and $V$ is the total system volume.
Essentially, \cref{eqn:SC} states that the combined effects of differential
sampling, preferential concentration and the sling effect can be represented by
the product $\langle w_r \rangle g(R)$.
A common association is that differential sampling and the sling effect
primarily affect the relative velocity statistics and therefore mainly enter
through $\langle w_r\rangle$, whereas preferential concentration manifests
itself in the RDF.
However, this separation is not strictly valid: caustics also contribute to
clustering~\cite{gustavssonStatisticalModelsSpatial2016}, and polydispersity
simultaneously increases
relative velocities while modulating preferential concentration, as we will
show in the present work.

Due to the massive separations of length scales of atmospheric flows,
precipitation must be parameterized in weather and climate models rather than
resolved directly.
Large-eddy simulations (LESs) with grid spacing of $\mathcal{O}(10\,\unit{m})$
have become the tool of choice to numerically investigate cloud droplet growth
and to develop parameterizations for coarser-resolution simulations.
Within LES frameworks, the collision kernel is central, because it directly
controls the conversion of small droplets into larger ones.
Typically, the kinematic form (\cref{eqn:SC}) is adopted, with $\langle
    w_r\rangle$ and $g(R)$ modeled separately based on droplet properties and local
flow conditions.
For example, two recent
studies~\cite{hoffmannRouteRaindropFormation2017a,chandrakarAreTurbulenceEffects2024}
have applied this approach showed that turbulence markedly broadens the droplet
size distribution in convective clouds.
Both relied on the celebrated kernel of Ayala et al.~\cite{ayalaEffectsTurbulenceGeometric2008a}.
However, the Ayala kernel was developed for droplets with settling parameters
$Sv>1$, where gravitational effects dominate over turbulent interactions.
Further developments are therefore needed to represent turbulence-enhanced
collisions across the full range of droplet sizes relevant for warm-rain
formation, and to account for the influence of turbulent intermittency.

\subsection{Onishi framework for the turbulent collision kernel}
\label{sec:onishi_model}
We proceed with a discussion of the Onishi framework for the turbulent collision
kernel, which serves as the main reference model in the comparisons that follow.
Originally proposed by Onishi et
al.~\cite{onishiLagrangianTrackingSimulation2015} and later refined by
Onishi~\&~Seifert~\cite{onishiReynoldsnumberDependenceTurbulence2016}, it
represents, to the best of our knowledge, the most comprehensive and advanced
model of the turbulent collision kernel currently available for
non-settling droplets.
The framework combines three separate models: an original formulation for the
monodisperse RDF, an expression for the bidisperse RDF following Zhou et al.~\cite{zhouModellingTurbulentCollision2001}, and the model for the RRV following Wang
et al.~\cite{wangStatisticalMechanicalDescription2000}.

The first ingredient is the monodisperse RDF.
Onishi~\&~Seifert~\cite{onishiReynoldsnumberDependenceTurbulence2016} propose
the empirically fitted form
\begin{equation}
    \label{eqn:onishi} g_{ii}(R) - 1 =
    \begin{cases}
        A_1 St^2 &
        \quad \text{for } St_i < St_a \\ A_2 Re_\lambda St^{-2} & \quad \text{for }
        St_a \leq St_i,
    \end{cases}
\end{equation}
where $A_1 = 110$, $A_2=0.38$ and
$St_a=((A_2/A_1)Re_\lambda)^{1/4}$ (the full formulation is given in Appendix~\ref{sec:onishi_appendix}).
The explicit dependence on $Re_\lambda$ is intended to capture the influence of
internal intermittency on preferential concentration at increasing Reynolds
numbers.
For $St_i<St_a$, the model predicts that clustering is essentially independent
of $Re_\lambda$, recovering the scaling
$g_{ii}(R)-1\varpropto St_i^2$, which is consistent with the leading-order
small-$St_i$ result of Wang et
al.~\cite{wangStatisticalMechanicalDescription2000}.

The second ingredient is the extension to droplets of unequal inertia, since
droplets with different Stokes numbers need not cluster in the same
regions of space.
Onishi et al. use the formulation of Zhou et al.~\cite{zhouModellingTurbulentCollision2001} 
to express the bidisperse RDF in terms of the monodisperse value of each species
as
\begin{equation}
    \label{eqn:gij_Zhou} g_{ij}(R) = 1 +
    \rho_{ij}[(g_{ii}(R)-1)(g_{jj}(R)-1)]^{1/2}.
\end{equation}
Here, $\rho_{ij}$
is a correlation coefficient quantifying the spatial overlap between the
droplet clusters of species $i$ and $j$.
By construction, $0\leq \rho_{ij}\leq 1$: values near unity correspond to
strong overlap, whereas $\rho_{ij}\to0$ indicates that the two clustering
patterns are effectively decorrelated.
The key assumption of Zhou et al. is that, to leading order, this overlap is
determined solely by the Stokes-number ratio and is symmetric under exchange of
the two species.
Defining
$\phi=\max(St_i/St_j,\,St_j/St_i)$, they therefore represent the correlation
coefficient by a single function $\rho(\phi)$.
Fitting separately in the ranges $1\leq \phi\leq 2.5$ and $\phi>2.5$ to DNS
data at $Re_\lambda=45$, and combining the resulting expressions with a
$\tanh$-based smoothing function, they obtain
\begin{equation}
    \label{eqn:rho_Zhou}
    \rho_{ij} = 2.6\exp(-\phi) + 0.205\exp(-0.0206\phi)
    \tfrac{1}{2}[1+\tanh(\phi-3)].
\end{equation}

Finally, Onishi and coauthors adopt the model of Wang et
al.~\cite{wangStatisticalMechanicalDescription2000} to evaluate the RRV.
In this framework, $\langle w_r \rangle$ is composed of a shear-induced and an
acceleration-induced contribution, $\langle w_r \rangle^2 = \tfrac{2}{\pi} (
    w^2_\text{shear} + w^2_\text{accel} ),$ a separation also adopted in more
recent models (see Ref.~\cite{ayalaEffectsTurbulenceGeometric2008a} for a review),
where differential sampling in polydisperse suspensions contributes primarily
through the acceleration term.
The model was formulated and validated at $Re_\lambda\leq 75$.
Explicit expressions for $w_\text{shear}$ and $w_\text{accel}$ following Wang
et al.~\cite{wangStatisticalMechanicalDescription2000} are given in
Appendix~\ref{sec:wang_appendix}.

\section{Methods}
\label{sec:methods}
We consider inertial particles suspended in three-dimensional homogeneous
isotropic turbulence.
The flow velocity field $\bm{u}(\bm{x},t)$ is the solution of the
incompressible Navier-Stokes equations.
We assume that particles are small ($r\ll\eta$, where $\eta=(\nu^3/\eps)^{1/4}$ is
the Kolmogorov length scale) and heavy compared to the carrier gas
($\rho_p/\rho_f=800$).
Under these conditions, particles neither disturb the flow nor experience
significant hydrodynamic interactions, so their motion can be described by
the point-particle approximation
\begin{equation}
    \label{eqn:pp} \dot{\bm{x}} =
    \bm{v}, \quad \dot{\bm{v}} = \frac{\bm{u}(\bm{x},t) - \bm{v}}{\tau_p} + \bm{g}.
\end{equation}
Here, $\bm{x}(t)$ and $\bm{v}(\bm{x}(t),t)$ are the Lagrangian particle
position and velocity, respectively, and
$\bm{g}=-g\bm{e}_z$
is the gravitational acceleration.
The point-particle approach is justified for atmospheric clouds, where typical
droplet volume fractions $\Phi_p = \tfrac{1}{V}\sum_{i=1}^{N_p}\tfrac{4}{3}\pi
    r_i^3$ are on the order of $\mathcal{O}(\Phi_p)=
    10^{-6}$~\cite{grabowskiGrowthCloudDroplets2013}.
Throughout this work, we will use the terms ``particle'' and ``droplet''
interchangeably.

We utilize isotropic homogeneous turbulent flow fields obtained from
an in-house solver ($Re_\lambda=55$ and $179$) as well as from
the Johns Hopkins Turbulence Database~\cite{liPublicTurbulenceDatabase2008} (JHTDB, $Re_\lambda=418$).
The fields are generated by solving the incompressible Navier-Stokes
equations using a pseudo-spectral method with low-wavenumber forcing.
In all cases, the computational domain consists of a periodic cube of side
length $L=2\pi$, discretized into $M$ grid points.
The flow characteristics and simulation parameters are given in~\cref{tab:turbparams}.
\begin{table}
    \caption{Summary of flow characteristics and computational parameters.}
    \label{tab:turbparams}
    \begin{ruledtabular}
        \begin{tabular}{cccccc}
            Case                                        & M      & $Re_\lambda$ & $\nu$      & $\eta$   & $\tau_\eta$ \\
            \hline
            In-house-256                                & $256$  & $55$         & $0.0075$   & $0.039$  & $0.2006$    \\
            In-house-512                                & $512$  & $179$        & $0.008$    & $0.007$  & $0.0613$    \\
            JHTDB~\cite{liPublicTurbulenceDatabase2008} & $1024$ & $418$        & $0.000185$ & $0.0028$ & $0.0424$    \\
        \end{tabular}
    \end{ruledtabular}
\end{table}

\Cref{eqn:pp} is solved for large ensembles of particles
by trilinearly interpolating the flow velocity to the particle positions and
using an integration scheme adapted from Jenny et al.~\cite{jennySolutionAlgorithmFluid2010}.
We use an in-house parallel solver designed to run on massively parallel
compute architectures.
The computational domain is decomposed into pencils, and
particles are distributed across \code{mpi}~\cite{mpi40} ranks.
Despite the point-particle approximation, a mass $m$ is assigned to every particle, thereby defining the radius
$r=(3m/(4\pi\rho_p))^{1/3}$ and, through~\cref{eqn:taup}, the relaxation time $\tau_p$.
Gravitational settling is neglected ($g=0$) in most of our simulations, except for those
discussed in \cref{sec:simulations_with_gravity,sec:dsd_broadening}.

Collision statistics are computed using the ghost-collision approach, where
particles pass through each other without interaction.
We determine the (dynamic) collision rate by counting the number of pairs per
unit time satisfying the contact condition $\left|\bm{\Delta x}\right| \leq R$.
To count efficiently, a binary tree data structure is constructed on each rank
during every time step, and particles close to boundaries are exchanged with
the neighboring ranks to check for inter-rank collisions.
Spherical shells of thickness $s = 0.1R$ are used to obtain consistent
results for the RRV and RDF (see
Ref.~\cite{wangStatisticalMechanicalDescription2000}), and the results are
averaged over ensembles of $N\geq10^8$ particles.
It was verified that the dynamic matches the kinematic collision
kernel within statistical uncertainty.

To obtain the statistics presented in \cref{sec:monodisperse_suspensions,sec:bidisperse_suspensions}, we initialize $N_p=5\cdot10^8$ particles per Stokes number uniformly at random
in the domain.
Precursor simulations are performed for a time $t_0=165\tau_\eta$ to reach
statistical stationarity of the collision rate.
The simulations are then restarted from $t_0$ and the ghost collision algorithm
is used to detect collisions.
When clustering is strong, colliding particles tend to stay together for
multiple time steps or come into contact again later in time.
To avoid overestimation of the collision rate in these cases (specifically for
monodisperse suspensions), the results collected during the first ten time
steps are discarded to exclude repeated contacts between particles that had
already collided at times $t < t_0$.
Each pair is only allowed to collide once.
Particle trajectories are linearly extrapolated to detect in-between-time step
collisions.

For the simulations with coalescence, we assume a collision efficiency of one.
This means that each colliding particle pair $(i,j)$ is replaced by a new
particle introduced at the position
$\bm{x}_{\text{new}}=(m_i\bm{x}_i+m_j\bm{x}_j)/(m_i+m_j)$ with
mass $m_{\text{new}} = m_i+m_j$ and velocity
$\bm{v}_{\text{new}}=(m_i\bm{v}_i+m_j\bm{v}_j)/(m_i+m_j)$.
The number of initialized particles varies across runs (as documented in
\cref{tab:params}) in order to keep the volume fraction $\Phi_p = 8.5\cdot 10^{-6}$
constant.
Particles that initially overlap with others are removed from the simulation.

\section{Results} \label{sec:results}
\subsection{Monodisperse suspensions}\label{sec:monodisperse_suspensions}
\begin{figure}[t]
    \includegraphics[width=0.3\linewidth]{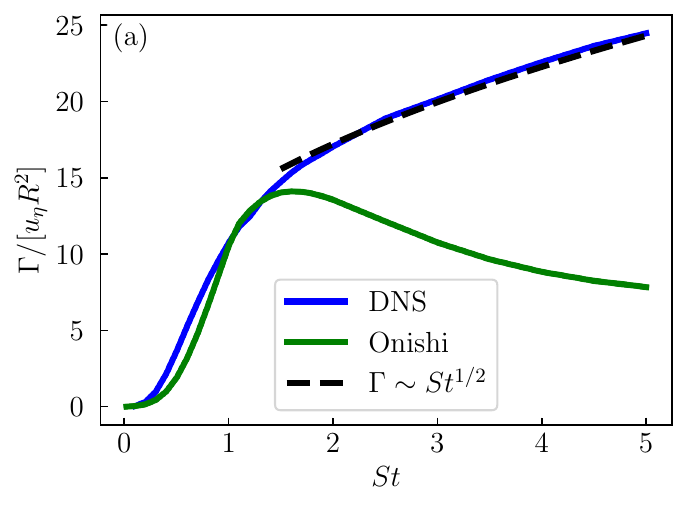}
    \includegraphics[width=0.3\linewidth]{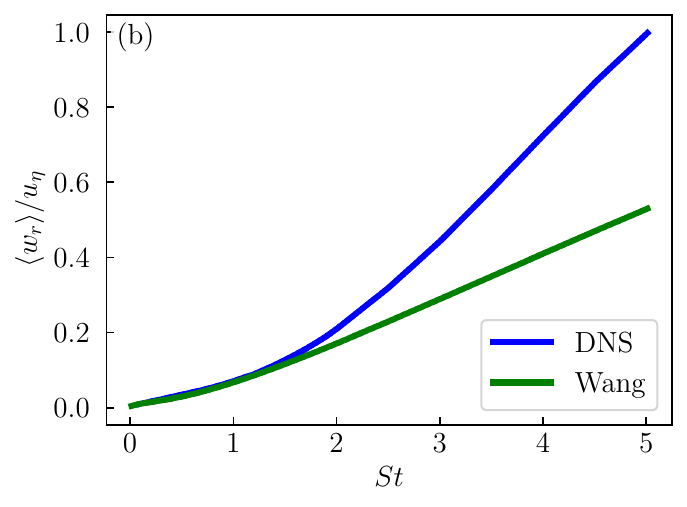}
    \includegraphics[width=0.3\linewidth]{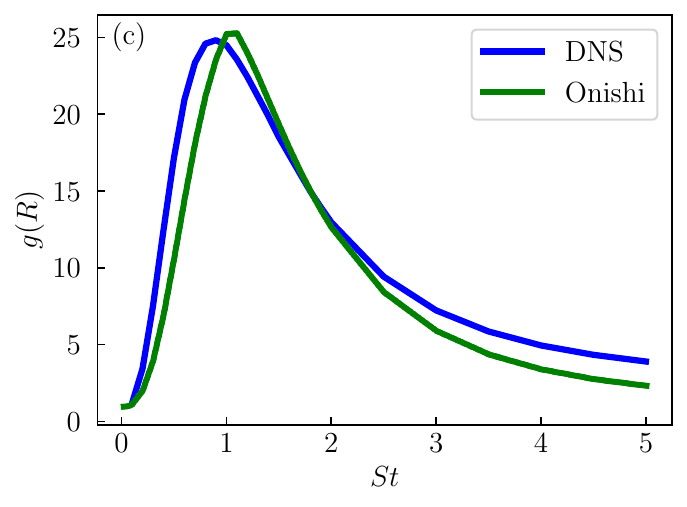}
    \caption{
        Comparison of monodisperse statistics at $Re_\lambda=418$:
        (a) collision kernel, (b) radial relative velocity (RRV) and (c) radial
        distribution function (RDF) at contact as a
        function of the Stokes number.
        DNS results (blue), are compared against model expressions (green).
        Note that the Wang model~\cite{wangStatisticalMechanicalDescription2000}
        was originally validated against $Re_\lambda\leq 75$ DNS data.
    }
    \label{fig:mono}
\end{figure}
The collision statistics for the monodisperse case are shown in \cref{fig:mono}.
In panel (a), the collision kernels obtained from the DNS (blue) and from the model
expressions (green) are plotted as a function of the Stokes number.
Following Refs.~\cite{vosskuhlePrevalenceSlingEffect2014,irelandEffectReynoldsNumber2016}, we normalize the collision kernel by $R^2u_\eta$.
Gravitational settling is neglected.
The collision kernel increases rapidly with the Stokes number, most notably in
the range $0.3 \leq St \leq 1$.
At larger $St$, the collision kernel scales as $\Gamma \sim St^{1/2}$, as shown
by the dashed black line.
The dependence of the collision kernel on the Stokes number can be understood
by examining the two terms appearing in \cref{eqn:SC}: the RRV and the RDF at
contact, shown in Figs.~\ref{fig:mono}(b) and (c), respectively.
The growth of $\langle w_r \rangle$ is gentle and approximately linear at
first, but increases more rapidly at larger $St$.
This behavior is attributed to the sling effect, which becomes increasingly
biases the relative velocity statistics between same-sized
particles at contact.
The RDF, on the other hand, is notably non-monotonic.
Particles disperse homogeneously in space with $St\to0$ and $St\to\infty$,
respectively, resulting in $g(R)=1$ in both limits.
In between, the RDF peaks at $St\approx1$, commonly identified as the Stokes
number with maximum clustering.
The dependence of $\Gamma$ on the $St$ can thence be interpreted as
the combined effect of $\langle w_r \rangle$ and $g(R)$.
It is tempting to attribute the steep increase of the collision kernel between
$St\approx0.3$ and $St\approx1$ solely to preferential concentration, since the
RDF grows at a slope far greater than that of the RRV.
However, separating preferential concentration and the sling effect in this
manner is not possible, since caustics are also involved in facilitating
clustering~\cite{gustavssonStatisticalModelsSpatial2016}.
A study by Vosskuhle et al.~\cite{vosskuhlePrevalenceSlingEffect2014}
showed that the sling effect already significantly contributes to the collision
dynamics at $St\gtrsim0.75$, yet the ensemble-averaged RRV only weakly affects
the product $\langle w_r \rangle g(R)$ in this range.

While preferential concentration decreases with Stokes numbers exceeding
$St=1$, the RRV continues to grow.
Beyond $St\gtrsim2$, our data suggest a scaling of the collision kernel
proportional to $St^{1/2}$.
This scaling is consistent with theoretical
predictions~\cite{wilkinsonCausticActivationRain2006,panRelativeVelocityInertial2010}
of the RRV in the sling-dominated regime at high Reynolds numbers, where
particle velocities become increasingly uncorrelated with both the
instantaneous flow field and with those of nearby particles.
In this regime, particle velocities are randomly distributed, making a
gas-kinetic approach following
Abrahamson~\cite{abrahamsonCollisionRatesSmall1975} more appropriate.
It has been suggested that the collision kernel should be expressed as the sum
$\Gamma = \Gamma_{\text{adv}} + \Gamma_{\text{sling}}$, explicitly
differentiating between the collision of particles that experience caustics and
those that do not~\cite{wilkinsonCausticActivationRain2006,vosskuhlePrevalenceSlingEffect2014}.

The Onishi model provides a remarkably good estimate for the monodisperse RDF,
especially considering that the model was used ``out-of-the-box'', without
re-fitting the parameters to our DNS data.
The Wang model, on the other hand, underestimates
$\langle w_r\rangle$ for $St\gtrsim 1.5$, i.e., it does not accurately capture
the enhanced relative motion of heavier particles due to the sling effect.
As a result, the combined expression for the monodisperse collision kernel
agrees well with the DNS data for $St\lesssim 1.5$, but fails to reproduce the
$\Gamma\sim St^{1/2}$ scaling at larger $St$.

\subsection{Bidisperse suspensions}\label{sec:bidisperse_suspensions}
\begin{figure}[t]
    \includegraphics[width=\linewidth]{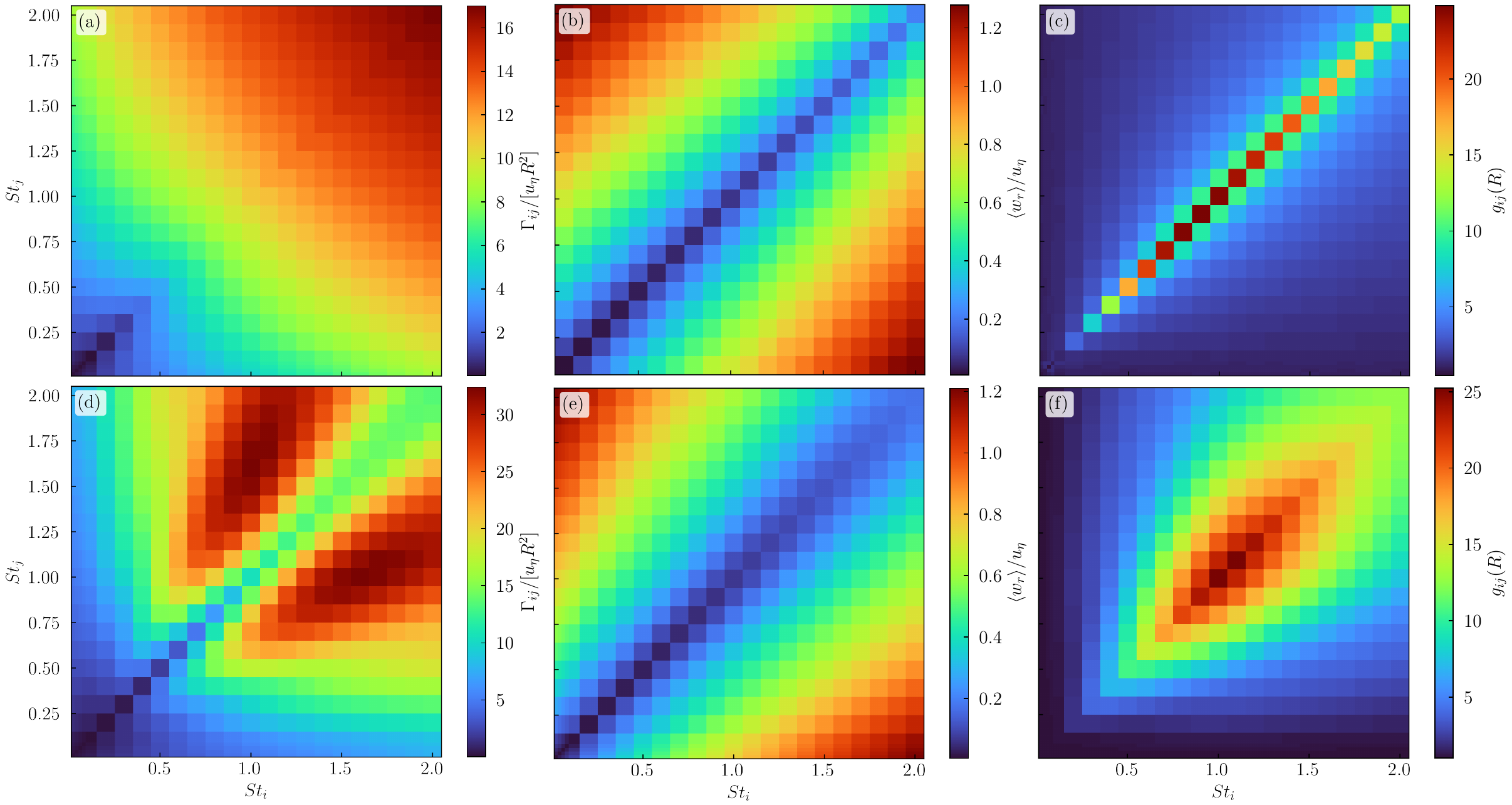} \caption{
        Bidisperse collision statistics for droplet pairs ($St_i,St_j$):
        (a-c) DNS results at $Re_\lambda=418$ (top row) compared with
        (d-f) predictions from Onishi/Wang model~\cite{onishiReynoldsnumberDependenceTurbulence2016,wangStatisticalMechanicalDescription2000} (bottom row).
        Columns represent the
        (a,d) collision kernel, (b,e) RRV and (c,f) RDF at contact, respectively.
    }
    \label{fig:bidisp}
\end{figure}

Now, how does polydispersity, naturally occurring in cloud droplet populations,
affect the collision dynamics?
To address this question, we compute the collision kernel, RRV and RDF at
contact for non-settling droplet pairs $(i,j)$ with Stokes numbers ranging from $St=0.02$ to
$St=2$.
The results are shown in the top row of \cref{fig:bidisp} as pseudocolor
matrices mapped across Stokes number pairs, while the bottom row shows the
corresponding model predictions.
The complete map of the collision kernel is provided in
\cref{tab:collision_map} for future reference.

In agreement with the monodisperse results, the collision rate generally
increases with the Stokes number, resulting from the combined effects of
$\langle w_r \rangle$ and $g(R)$.
We observe that the RRV increases with the Stokes number separation within
the colliding pair, and that $\langle w_r \rangle_{ij} \geq
\max(\langle w_r\rangle_{ii},\langle w_r\rangle_{jj})$.
The enhancement of the RRV due to differential sampling largely outweighs
its increase along the diagonal.
Conversely, the RDF peaks along the diagonal and quickly drops as the
Stokes numbers separate.
This is indicative of the rapid decorrelation of the particle clusters due to
differential sampling.

We observe that the behavior of the bidisperse kernel falls into two regimes.
At small Stokes numbers (i.e., when both $St_i, St_j \lesssim 0.4$),
polydispersity enhances collisions even when it reduces the Stokes
number of one of the colliding droplets.
This behavior is evident in \cref{fig:bidisp}(a): starting from a point on the
diagonal ($St_i, St_j \leq 0.4$), moving horizontally to the left or vertically
downward--thus decreasing one Stokes number while keeping the other
fixed--results in an increase in $\Gamma$.
To explain this behavior, consider the competing effects of polydispersity to
decrease particle clustering yet increase relative motion.
At small Stokes numbers, particles tend not to detach from the flow, and
preferential concentration is weak.
As a result, the enhancement of the relative velocities due to differential
sampling is more significant, and the net effect of polydispersity on the
collision kernel is positive.
Conversely, at larger Stokes numbers, $St \gtrsim 0.4$, preferential
concentration significantly contributes to the collision rate, and the loss of
clustering due to polydispersity outweighs the increase in relative velocities,
leading to an attenuation of collisions.
In this regime, the collision kernel increases only when one of the Stokes
numbers--and thus the overall inertia of the droplet pair--becomes larger.

James \& Ray~\cite{jamesEnhancedDropletCollision2017} have reported bidisperse
collision kernels obtained from DNS of two-dimensional particle-laden
turbulence.
Their data show a pronounced increase in the collision rate along the diagonal,
indicating that collisions between same-sized particles are strongly enhanced.
Our results do not exhibit this behavior; instead, the collision kernel varies
smoothly with the Stokes number separation.
In preparing this paper, we identified that this discrepancy may arise from the
algorithm used to detect collisions.
In monodisperse suspensions, colliding droplet pairs can remain in contact
over prolonged periods of time or re-enter contact later, leading to an
overestimation of the collision rate, particularly under conditions of strong
clustering.
To mitigate this effect, the first ten time steps of the detection procedure
are discarded, and each droplet pair is allowed to collide only once, as
described in \cref{sec:methods}.
More fundamentally, the use of two-dimensional turbulence limits the
applicability of their results to atmospheric conditions.
The distinct characteristics of 2D turbulence, such as the inverse energy
cascade, may not accurately capture the collision dynamics relevant to cloud
droplets.
Two-dimensional turbulence lacks the flow structures and intermittency
associated with three-dimensional turbulence, which are possibly crucial for
the collision dynamics of cloud droplets.
For example, it has been
demonstrated~\cite{meibohmCausticFormationNonGaussian2024,codispotiDissectingInertialClustering2025}
that caustic formation occurs through specific flow topologies specific to
three-dimensional turbulence.
Moreover, Onishi \& Vassilicos~\cite{onishiCollisionStatisticsInertial2014}
compared the collision statistics in 2D and 3D DNS and found that inertial
clustering is reduced with increasing $Re_\lambda$ due to internal
intermittency, absent in 2D turbulence.

\Cref{fig:bidisp}(d) shows the collision kernel obtained from the Onishi model,
with the corresponding expressions for $\langle w_r\rangle$ (Wang model, see \cref{sec:wang_appendix})
and $g_{ij}(R)$ (\cref{eqn:onishi,eqn:gij_Zhou,eqn:rho_Zhou}) shown in Figs.~\ref{fig:bidisp}(e) and (f), respectively.
We find that the Wang model captures the qualitative trends of the bidisperse
RRV.
For $St<1.0$, $\langle w_r\rangle$ agrees with the DNS results to within 10\%.
Larger errors are observed mainly for differently-sized particles at larger
Stokes numbers, for which the Wang model underestimates the relative velocities
due to differential sampling.
The bidisperse RDF, on the other hand, is not estimated accurately by the Zhou
model.
The DNS data display a more rapid decorrelation of the droplet clusters
compared to the predictions provided by \cref{eqn:rho_Zhou}.
As a result, the Onishi model strongly overestimates the bidisperse collision
kernel, specifically for droplet pairs with Stokes numbers $St_i,\,St_j>1$.
The relative error with respect to the DNS data of the Onishi model is shown in
\cref{fig:model_and_errors}(d).

\subsubsection{Bidisperse correlation coefficient}
\label{sec:bidisperse_correlation_coefficient}
\begin{figure}
    \includegraphics[width=0.35\linewidth]{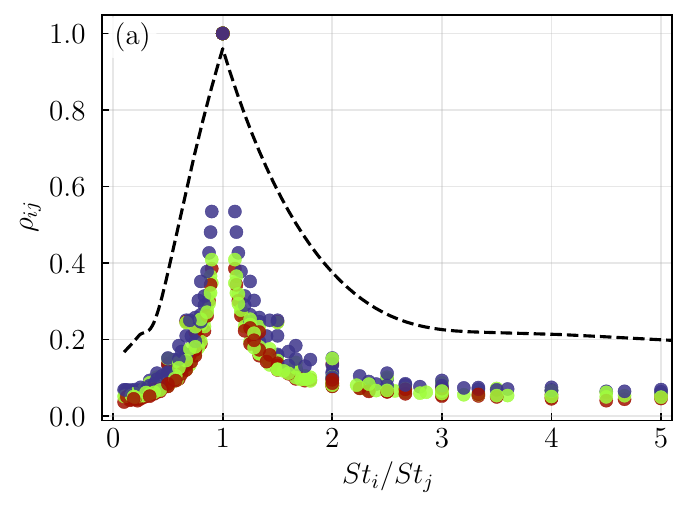}
    \includegraphics[width=0.35\linewidth]{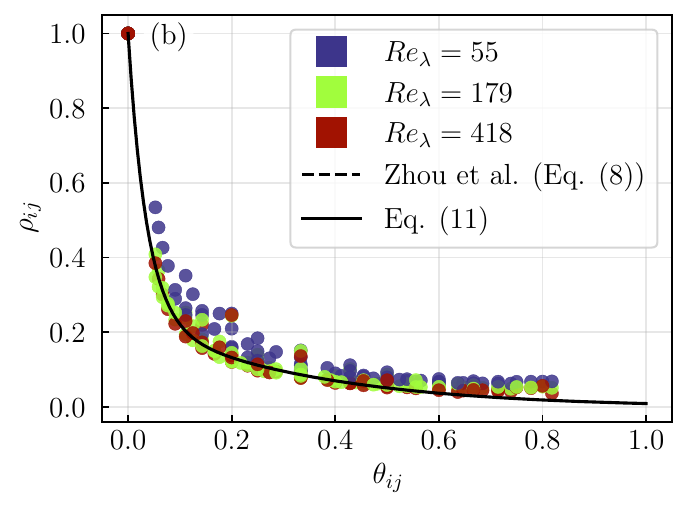}
    \caption{
        Evaluation of bidisperse clustering:
        (a) Cross-species correlation coefficient versus Stokes number ratio
        obtained from the DNS at $Re_\lambda = 55$, $179$ and $418$
        ($St\in[0.2,2.0]$)
        and with the Zhou model (\cref{eqn:rho_Zhou}, originally validated at $Re_\lambda=45$, black dashed line).
        (b) The same data plotted against the dispersion parameter, $\theta_{ij}$,
        together with the prediction from~\cref{eqn:rho_theta} (black solid line).
    }
    \label{fig:rho}
\end{figure}

Comparing Figs.~\ref{fig:bidisp}(c) and (f) shows that the Zhou model~\cite{zhouModellingTurbulentCollision2001}
overestimates bidisperse clustering, indicating that the correlation
coefficient $\rho(\phi)$ prescribed by \cref{eqn:rho_Zhou} decays too slowly
with increasing $\phi = \max(St_i/St_j,St_j/St_i)$.
Indeed, we find that the droplet clusters decorrelate faster than the model
suggests.
This is apparent in \cref{fig:rho}(a), which shows
the correlation coefficient $\rho_{ij}=(g_{ij}(R) -
    1)/[(g_{ii}(R)-1)(g_{jj}(R)-1)]^{1/2}$ obtained from the
DNS as a function of $St_i/St_j$ alongside the prediction from \cref{eqn:rho_Zhou}.
Here, we show droplet pairs with Stokes numbers in the range
$St\in[0.2,2.0]$, since clustering is very weak for $St\geq0.2$, even in the 
monodisperse case.
We find that our data collapse quite well, for all simulated Reynolds numbers.
The Zhou model, however, exhibits systematic discrepancies relative to the DNS data.
In addition, we find that the correlation coefficient is not quite symmetric around
$St_i/St_j=1$, suggesting that the assumption fundamental to the derivation of Zhou et
al.~\cite{zhouModellingTurbulentCollision2001} is physically unjustified.

We now provide arguments to explain the observed behavior.
First, we ask why the Zhou model systematically overestimates the bidisperse
overlap.
Ayala et al.~\cite{ayalaEffectsTurbulenceGeometric2008a} observed similar errors but
concluded they were a consequence of neglecting gravity in the model's derivation.
In contrast, the present work demonstrates that the model remains inaccurate
even for non-settling droplets.
Since the Zhou model was derived at a relatively low $Re_\lambda=45$, it is
tempting to assume that the observed discrepancies stem from the larger Reynolds
numbers used in this work.
Somewhat unexpectedly, however, our data for all three simulated $Re_\lambda = 55$,
$179$ and $418$ collapse quite well, suggesting that $Re_\lambda$
is not the primary driver of the error.
While the lowest Reynolds number case ($Re_\lambda=55$) deviates slightly from
the trend--potentially because a narrow inertial subrange may cause clustering
length scales to overlap with energy-containing scales--this remains a minor effect.
Instead, we propose that the discrepancy may be an artifact of the ``frozen''
flow field (i.e., advecting particles through a static velocity field) used
in the original derivation.
A previous study~\cite{zhouCollisionRateSmall1998} indicated that this
approach can lead to an overestimation of particle clustering.
While we do not verify this directly here, it stands as the most
plausible explanation for the model's systematic failure to reproduce the
current DNS results.

Note that the cross-species correlation coefficient is symmetric, i.e.,
$\rho(St_i,St_j)=\rho(St_j,St_i)$.
The apparent asymmetry around $St_i/St_j=1$ in \cref{fig:rho}(a) arises because
$\rho$ is plotted as a function of $St_i/St_j$ while $St_j$ is held fixed.
For $St_i<1$, decreasing $St_i$ approaches the tracer limit, leading to a
faster decay of the cross-correlation.
The asymmetry $\rho(St_i/St_j)\neq\rho(St_j/St_i)$ motivates the use of an
alternative parameter to replace $\phi = \max(St_i/St_j,St_j/St_i)$ in modeling
the correlation coefficient.
The dispersion parameter
\begin{equation} \label{eqn:theta}
    \theta_{ij} =
    \frac{St_i-St_j}{St_i+St_j}
\end{equation}
provides a suitable alternative, as
the collision statistics are naturally symmetric with
$\theta_{ij}=-\theta_{ji}$, so that the modeling task reduces to finding an
expression for $\rho(\theta)$, where $\theta=|\theta_{ij}|$.
Moreover, the dispersion parameter offers a quantification of the
$St$-separation \emph{relative} to the overall inertia of the droplet pair.
Based on this reasoning, we expect $\rho(\theta)$ to exhibit a more universal
behavior compared to $\rho(\phi)$, which \cref{fig:rho}(b) confirms in the
affirmative.
The majority of data points collapse on an exponential curve, so that
\begin{equation}
    \label{eqn:rho_theta} \rho(\theta) = 0.77  \exp(-27.9\theta)
    + 1.96 \exp(-4.67\theta)\tfrac{1}{2}[1+\tanh(\theta-1)]
\end{equation}
provides an excellent fit to the DNS data.
To obtain this expression, we used the composite formulation from the Zhou
model and performed a least-squares fit on the $Re_\lambda=179$ and $418$ DNS data.
In general, we conclude that our formulation improves the predictions of the
bidisperse correlation coefficient in the absence of gravity, while exhibiting
negligible sensitivity to $Re_\lambda$.

\subsubsection{A novel parameterization for the bidisperse collision kernel}
\label{sec:parameterization}
\begin{figure}[t]
    \includegraphics[width=\linewidth]{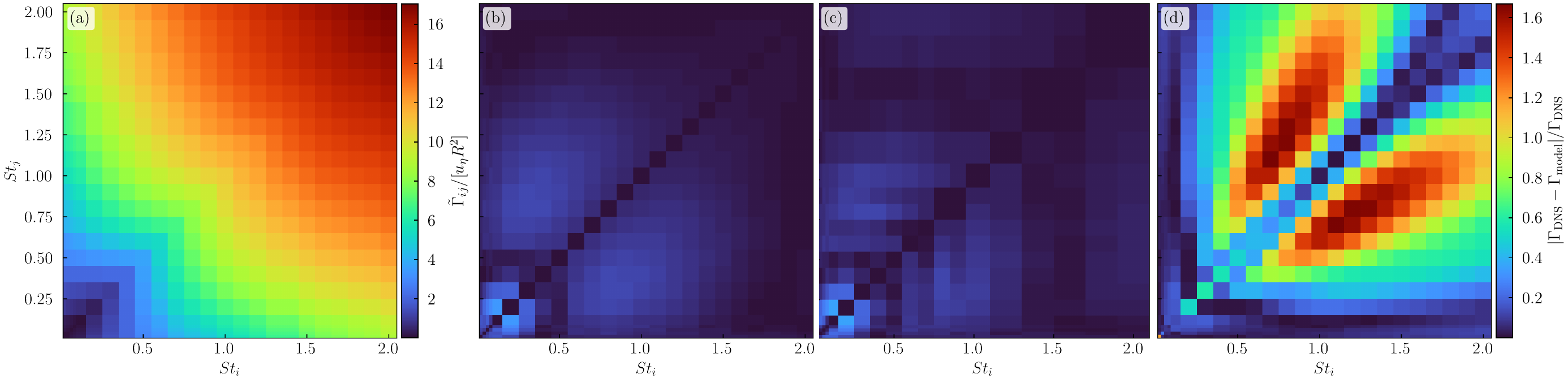}
    \caption{
        Validation of the proposed parameterization (\cref{eqn:parameterization}):
        (a) Predicted kernel matrix using DNS-derived $\Gamma_{ii}$ and $\Gamma_{0i}$
        ($St_0=0.02$) at $Re_\lambda=418$.
        Relative error of the parameterization with respect to the DNS at (b)
        $Re_\lambda=418$
        and (c) $Re_\lambda=179$, compared to (d) the error obtained from the
        Onishi/Zhou model.
        A common color scale is used for panels (b-d).
    }
    \label{fig:model_and_errors}
\end{figure}

In the previous section, we showed that although the Onishi model accurately predicts
monodisperse clustering, it produces large errors in the bidisperse case, as
evident from \cref{fig:model_and_errors}(d).
This limitation arises from its reliance on the Zhou model, which fails to
capture the rapid decorrelation of particle clusters as $St_i$ and $St_j$
diverge.
Having introduced an improved version of this model (\cref{eqn:rho_theta}), we
now move on to present a compact parameterization for the bidisperse collision
kernel as a function of the dispersion parameter, $\theta$.
This is motivated by the observation that the kernel varies smoothly as the
Stokes numbers separate, whereas both the RDF and RRV do not.
Letting $\tau = St_i + St_j$ and $\delta = St_i - St_j$, we consider the first-order
approximation $\Gamma(\tau,\delta) = \Gamma(\tau,0) + E(\tau)\tfrac{\delta}{\tau} +
    \mathcal{O}(\delta^2)$.
Here, the sign and magnitude of the polydisperse ``enhancement factor'' $E(\tau)$
depend on the overall inertia $\tau$, but not on the separation $\delta$.
As established earlier, the effect of polydispersity arises from the
competition between increased relative velocities and reduced preferential
concentration as $St_i$ and $St_j$ diverge.
To quantify this balance, we seek a reference quantity that isolates the
velocity enhancement from spatial correlations.
On the basis of empirical observations, we find that the kernel
$\Gamma_{0i}$, describing interactions between tracer-like droplets
($St_0 \ll 1$) and droplets with $St_i$, provides a suitable measure, as it
primarily captures the contribution from differential sampling.
Hence, the net enhancement due to polydispersity is approximated by the
difference $E(\tau)=\Gamma_{0i}-\Gamma_{ii}$.

\begin{figure}[t]
    \includegraphics[width = 0.37125\linewidth]{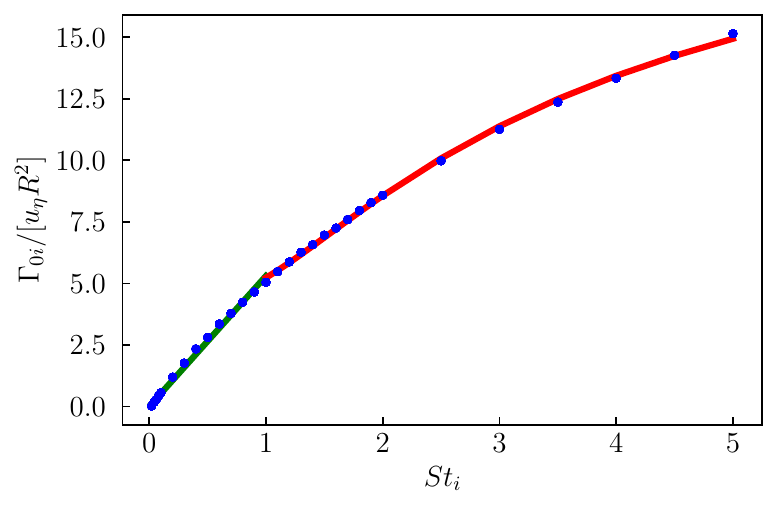}
    \caption{$\Gamma_{0i}$ obtained from the DNS at $Re_\lambda=418$ with
        $St_0=0.02$ (blue dots).
        The green and red lines depict the empirical fits $\Gamma_{0i}=aSt_i$ with
        $a=5.3$ and $\Gamma_{0i}=b_0+b_1\exp(-b_2/St_i)$ with
        $b_0=4.3,~b_1=19.6,~b_2=3.05$, respectively.
    }
    \label{fig:Gammai0}
\end{figure}

Following this approach, we find that the bidisperse collision kernel for a
droplet pair $(i,j)$ with $St_i \geq St_j$ can be effectively parameterized as
\begin{equation}
    \label{eqn:parameterization} 
    \tilde{\Gamma}_{ij} = \theta_{ij}\Gamma_{0i} + (1-\theta_{ij})\Gamma_{ii}.
\end{equation}
Here, $\Gamma_{ii}$ is the monodisperse collision kernel and $\Gamma_{0i}$ is
the collision kernel between particles of $St_i$ and very light,
close-to-tracer particles (shown in \cref{fig:Gammai0} with $St_0=0.02$).
Note that $\Gamma_{ii}$ corresponds to the diagonal and $\Gamma_{0i}$ to the bottom
row (or first column) of \cref{fig:bidisp}(a).
The parameterization is given for $St_i \geq St_j$ so that $\theta_{ij} \geq
    0$, but extension to arbitrary pairs follows directly from symmetry.
The heuristic behind expression~\eqref{eqn:parameterization} is the separation
of the collision kernel into inertial effects (preferential concentration and
the sling effect, represented by the monodisperse kernel) and differential
sampling (represented by $\Gamma_{0i}$).
The resulting bidisperse kernel blends the two contributions, weighted
by the dispersion parameter $\theta_{ij}$.
One easily verifies that as $St_j\rightarrow0$, the correct asymptotic behavior
$\tilde{\Gamma}_{ij}=\Gamma_{0i}$ is recovered.
Similarly, for $i=j$, \cref{eqn:parameterization} simplifies to the
monodisperse collision kernel.

\Cref{fig:model_and_errors}(a) displays the predictions of \cref{eqn:parameterization},
allowing direct comparison to the DNS reference (\cref{fig:bidisp}(a)).
The compact linear formulation captures the effect of differential sampling on
the collision kernel with remarkable qualitative accuracy.
It reproduces the enhancement due to polydispersity at small Stokes numbers,
even when the mean Stokes number of the colliding pair is reduced.
This behavior arises because initially, $\Gamma_{0i}$ increases more rapidly
with $St_i$ through differential sampling than $\Gamma_{ii}$ increases through
inertial effects, leading to $\Gamma_{0i}>\Gamma_{ii}$ for small $St_i$ and
$\Gamma_{ii}>\Gamma_{0i}$ for $St_i \gtrsim 0.4$.

Figures~\ref{fig:model_and_errors}(b,c) show the relative errors of our model
with respect to the DNS data at $Re_\lambda=418$ and $Re_\lambda=179$, respectively.
For both Reynolds numbers, the predictions generally agree with the DNS results
to within 10\%.
Larger errors are seen primarily for small Stokes numbers around the diagonal,
since the collision kernel is near zero in this region, thus amplifying the
relative error signal.
Overall, the predictions of our parameterization fit the DNS data significantly
better than the Onishi/Zhou model, whose relative error is shown in
\cref{fig:model_and_errors}(d).
The Onishi/Zhou predictions deviate most strongly for off-diagonal pairs
with $St\gtrsim1$, where the overestimation of the bidisperse
correlation coefficient with \cref{eqn:rho_Zhou} is most pronounced.

Recall that $\Gamma_{0i}$ describes the collisions between droplets with
$St_0\ll1$ (tracer-like) and $St_i$.
In order for \cref{eqn:parameterization} to yield a complete model for the
bidisperse collision kernel, an explicit expression for $\Gamma_{0i}$ is
required (whereas $\Gamma_{ii}$ can be modeled with the Onishi model).
In this regime, preferential concentration is negligible
($g_{i0}(R)\approx1$ for sufficiently small $St_0$), and the collision dynamics
are dominated by the RRV.
James \& Ray~\cite{jamesEnhancedDropletCollision2017} derived analytical
expressions for $\langle w_r\rangle$ under various limiting assumptions.
They demonstrate that with $St_j \ll1$, the RRV obeys the linear scaling
$w_r\sim St_i$ if $St_i \lesssim 1$ and exponential scaling of the form $w_r
    \sim \exp(-1/St_i)$ for $St_i \gg 1$.
Since $g(R)\approx 1$ in this regime, we can translate the scaling of the RRV to
the collision kernel.
In \cref{fig:Gammai0}, we show the $\Gamma_{0i}$ kernel obtained using
$St_0=0.02$ and apply the two scaling laws.
The DNS data are well described by these fits.

\subsubsection{Influence of gravitational settling}
\label{sec:simulations_with_gravity}
Because this work is primarily concerned with the role of turbulence in
enhancing collisions between cloud droplets, gravity has been neglected in the
simulations discussed thus far.
While these results have specific relevance to the warm rain context--particularly
for small droplets ($r\approx 10 \mum$) that encounter intermittent
high-dissipation events--their direct applicability to the onset of
precipitation is limited.
Under mean-field conditions, and as droplets grow beyond
$r\approx 15\,\mum$, gravitational settling becomes an important
factor in the collision-coalescence process~\cite{grabowskiGrowthCloudDroplets2013,vaillancourtReviewParticleTurbulenceInteractions2000}.
To assess the robustness of our parameterization for settling droplets, we
performed additional DNS including gravity.
A uniform gravitational acceleration $g$ is applied along the negative $z$-direction.
The Froude number is set to $Fr = \eps^{3/4} / (\nu^{1/4}g) = 0.173$,
representative of a cumulus cloud with $\eps = 0.05\,\unit{m^2/s^3}$.
With Stokes numbers spanning $St \in [0.02, 2]$, the settling
parameters range from $Sv = St/Fr \approx 0.12$ to $11.6$.
Note that this corresponds to cloud droplet radii of approximately $r\in[5.4,54]\,\mum$.

\begin{figure}[t]
    \includegraphics[width=0.33614\linewidth]{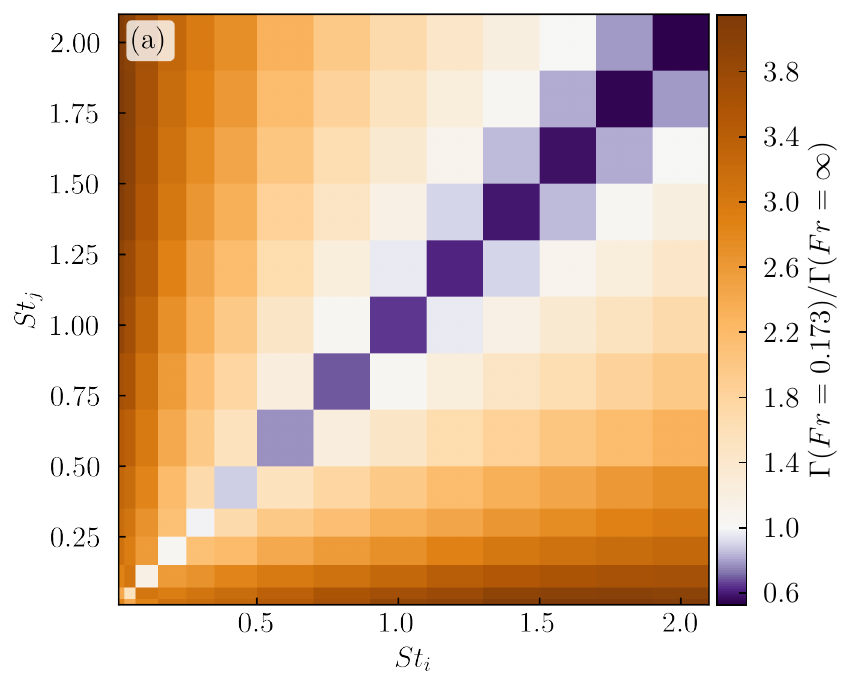}
    \includegraphics[width=0.33614\linewidth]{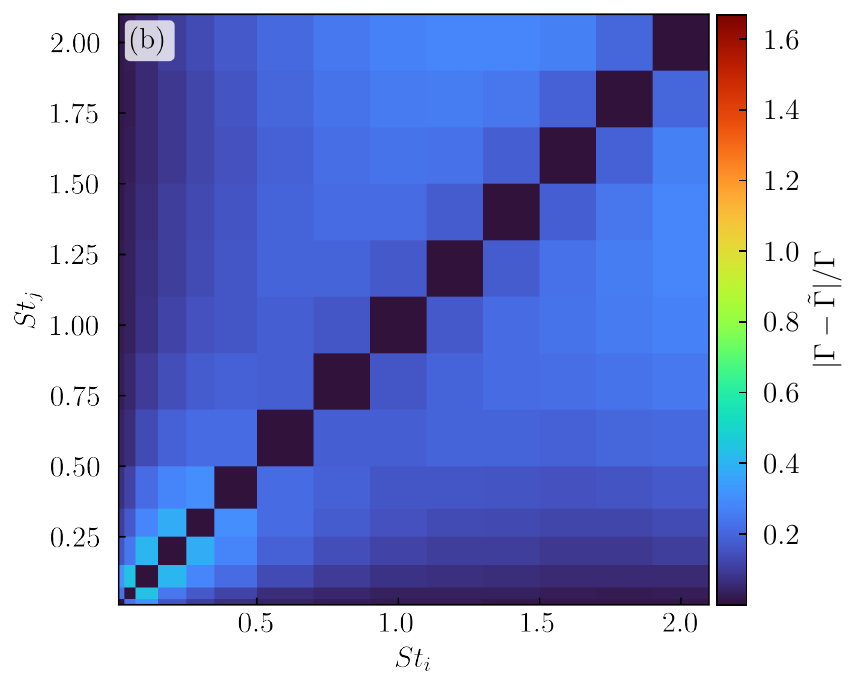}
    \caption{
        Influence of gravity on bidisperse collisions:
        (a) Ratio between settling ($Fr=0.173$) and non-settling collision kernels
        and
        (b) relative errors incurred by our parameterization (\cref{eqn:parameterization}).
        Note that the color scale of panel (a) is asymmetric.
        The color scale in panel (b) is identical to that used in Figs.~\ref{fig:model_and_errors}(b-d).
    }
    \label{fig:gravity}
\end{figure}

\Cref{fig:gravity}(a) displays the ratio between the settling and non-settling
collision kernel, mapped across Stokes number pairs.
In this figure, the droplet radius (and thus $Sv$) is varied, but we use the
Stokes number to keep the visualizations consistent with the preceding sections.
Collisions between same-sized droplets (along the diagonal) are attenuated by
settling.
This result is consistent with earlier monodisperse
studies~\cite{irelandEffectReynoldsNumber2016a}.
Physically, gravity advects droplets through the coherent flow
structures on timescales shorter than the local eddy turnover time, thereby
reducing the residence time available for accumulation in strain-dominated
regions.
This weakened interaction between the droplets and the flow manifests itself in
attenuated preferential concentration and fewer sling events.
As a result, both the RDF and the RRV are reduced for monodisperse droplets
compared to the non-settling case, and the collision rate is lower.
Away from the diagonal, however, the picture reverses.
Droplets of different sizes settle at different terminal velocities, adding a
systematic contribution to the relative velocity that grows with size separation.
This mechanism strongly enhances collisions, so that the off-diagonal entries of
the kernel matrix increase substantially relative to the zero-gravity case.

A key question is whether the parameterization for the bidisperse collision
kernel derived in the absence of
gravity (\cref{eqn:parameterization}) remains valid for settling droplets,
for which differential settling is the dominant collision mechanism at large
size separations.
Our data suggest that it does:
\Cref{fig:gravity}(b) shows the relative error of our model against the
DNS at $Fr=0.173$.
The errors are generally larger compared to the non-settling case, yet the
parameterization continues to reproduce the full bidisperse kernel
qualitatively across the majority of the parameter space.
This confirms that the underlying structure of the model--a linear blend between
the monodisperse and polydisperse effects weighted by $\theta$--captures
the essential physics even when differential settling is the
dominant source of relative motion.
In the presence of gravity, we find that $\Gamma_{0i} > \Gamma_{ii}$ for all $St_i$,
implying that the polydisperse contribution consistently exceeds the
monodisperse one.

\subsection{Turbulent broadening of the droplet size distribution} \label{sec:dsd_broadening}
We now shift from the ghost collision approach to simulations involving
coalescing droplets.
These simulations allow us to study the broadening of the droplet
size distribution (DSD), without relying on the simplifying assumptions of
Smoluchowski-type collision-coalescence models, which underestimate the
rapid formation of lucky droplets through successive correlated
collisions~\cite{becAbruptGrowthLarge2016}.
As discussed in the Introduction, the inertia and settling of cloud droplets vary
along their trajectories because the local turbulence intensity fluctuates in
space and time.
Consequently, collision-coalescence is modulated by the
instantaneous local dissipation rate.
To study the impact of $\eps$ on droplet growth without the computational
expense of additional simulations, we exploit the
scaling relations $St \varpropto r^2\eps^{1/2}$ and $Sv\varpropto r^2\eps^{-1/4}$.
In particular, increasing the dissipation rate by a factor $\alpha^4$ affects
droplet inertia in a manner equivalent to scaling the droplet
radii by $\alpha$ while keeping the flow field unchanged (whereas the setting
parameter decreases by $1/\alpha$).
% Since $St\varpropto r^2 \eps^{1/2}$, increasing the dissipation rate by a factor $\alpha^4$ affects droplet inertia in a manner equivalent to scaling the droplet
% radii by $\alpha$ while keeping the flow field unchanged.
This equivalence enables us to emulate dissipation rate fluctuations by
appropriately scaling the initial DSD.

We emphasize that the simulations are performed at fixed
$Re_\lambda=418$, well below the realistic values in atmospheric clouds
($\mathcal{O}(10^4)$).
As noted in the Introduction, clustering shows a weak, yet non-negligible
$Re_\lambda$-sensitivity, since extrapolation of the DNS to atmospheric
conditions spans at least one order of magnitude.
On the basis of the cited
reports~\cite{onishiLagrangianTrackingSimulation2015,onishiReynoldsnumberDependenceTurbulence2016,irelandEffectReynoldsNumber2016},
we therefore expect our simulations to slightly overestimate the growth of
droplets with $0.4\lesssim St\lesssim1$.
At higher $Re_\lambda$, dissipation rate fluctuations become more intense yet
more localized in space and time.
While the aim of our study is to demonstrate accelerated droplet growth in
highly dissipative parcels, by no means do we claim that our simulations
reproduce the cloud-realistic statistics of $\eps$.

Droplet populations are initialized with radii following a Gaussian
distribution, $r \sim \mathsf{N}(r_0, \sigma_r^2)$, with mean $r_0$ and
standard deviation $\sigma_r$, as is expected to result from condensation.
Both $r_0$ and $\sigma_r$ are scaled by~$\alpha$, so that a shift in the
local energy dissipation rate by a factor of $\eps/\eps_0=\alpha^4$ is emulated.
For example, by setting $\alpha=1.23$, we adjust the mean Stokes number from
$\langle St\rangle= 0.15$ to $\langle St\rangle = 0.23$, corresponding to a
$2.28$-fold increase in the local energy dissipation rate.
The parameters of all cases are given in \cref{tab:params}.
Since the droplets grow upon collision, gravitational settling must be accounted
for in these simulations, and the Froude number scales with $\alpha^3$.
To keep the droplet volume fraction $\Phi_p \varpropto n r^3$ unchanged, the
initial number of droplets, $N_p(t=0)$, is reduced as $\alpha$ grows.
A cloud-realistic value of $\Phi_p = 8.5\cdot 10^{-6}$ is kept constant
throughout the simulation campaign.

\begin{table}[t]
    \caption{
        Computational parameters for the simulations with coalescing droplets:
        Scaling factor $\alpha$, corresponding increase in the dissipation rate,
        Froude number, mean Stokes number, mean settling parameter, and initial
        droplet count.
    }
    \label{tab:params}
    \begin{ruledtabular}
        \begin{tabular}{cccccc}
            $\alpha$ & $\eps/\eps_0$ & $Fr$ & $ \langle St \rangle $   & $ \langle Sv \rangle $ & $N_p(t=0)$     \\
            \hline
            $1$    & $1   $ & $0.173$ & $0.15 $ & $0.87$ & $9.3\cdot10^8$ \\
            $1.23$ & $2.28$ & $0.322$ & $0.23 $ & $0.71$ & $5\cdot10^8$   \\
            $1.63$ & $7.06$  & $0.749$ & $0.40 $ & $0.53$ & $2.1\cdot10^8$ \\
            $2.15$ & $21.4$ & $1.719$ & $0.69 $ & $0.38$ & $9.3\cdot10^7$ \\
        \end{tabular}
    \end{ruledtabular}
\end{table}

\begin{figure}[t]
    \includegraphics[width=0.33614\linewidth]{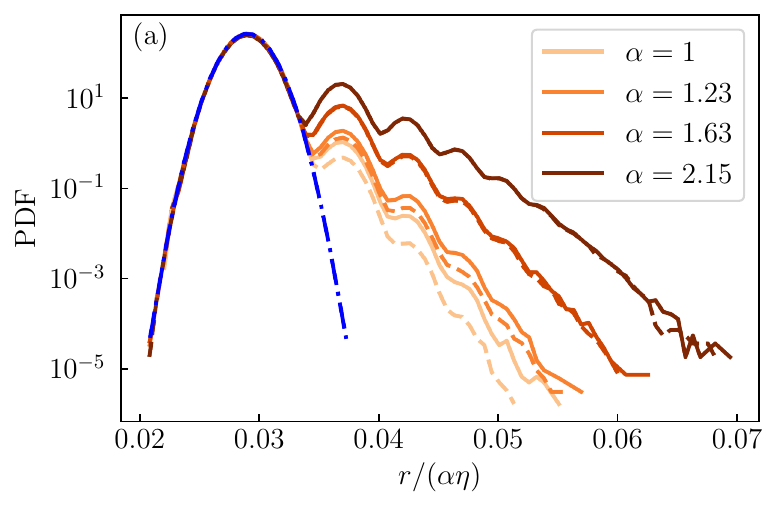}
    \includegraphics[width=0.27\linewidth]{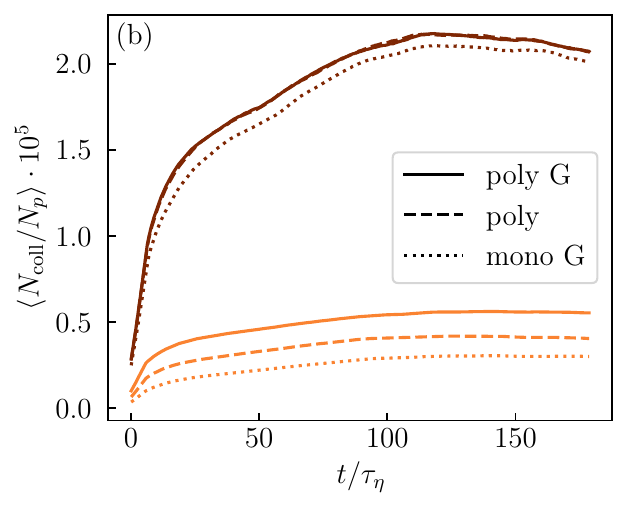}
    \caption{
        DNS results for coalescing droplets.
        (a) Probability density functions of the
        droplet radii at time $t=340\tau_p$ for different
        values of $\alpha$.
        The DSD at $t=0$ is indicated by the blue dash-dotted line;
        dashed lines provide a comparison to reference cases without gravity.
        (b) Mean collision-coalescence count (number of collisions
        normalized by the total number of droplets) versus time for $\alpha = 1.23$ and
        $2.15$.
        Solid lines denote the standard case with gravity, initialized with
        polydisperse droplet populations.
        For comparison, results are shown for reference simulations without
        gravity (dashed lines) and for a monodisperse case with gravity (dotted lines).
        The data are ensemble-averaged across all droplets and smoothed over
        intervals of $10\tau_\eta$.
    }
    \label{fig:dsdevolution}
\end{figure}

\Cref{fig:dsdevolution}(a) shows the DSDs obtained for different values of
$\alpha$ after an integration time of $t=340\tau_p$.
Since we emulate dissipation fluctuations by altering the droplet phase rather
than the flow field, the evaluation time is fixed relative to $\tau_p$ rather
than $\tau_\eta$.
The dashed-dotted blue line shows the initial DSD, whereas the dashed lines
show the equivalent DSDs obtained from reference runs without gravity.

All simulations are carried out in non-dimensional units.
To relate these results to a physical cumulus cloud, we note that a
representative Kolmogorov length scale is $\eta \approx 0.4\,\unit{mm}$~\cite{siebertHighresolutionMeasurementCloud2015}.
In this context, the initial droplet size corresponds to approximately
$r_0 \approx 11.6\,\mum$, and all droplets initially have $r < 15\,\mum$.
Despite these small radii, the droplets effectively collide and coalesce in the
turbulent flow within the short
simulated time interval, even in the base case ($\alpha = 1$).
The DSD broadens rapidly from its initial narrow Gaussian
distribution, developing long tails with visibly ``bumpy'' profiles, owing to the
narrow initial size spread combined with the discrete, binary nature of
coalescence events.
The long tails indicate that a few lucky droplets have grown disproportionately
(note the logarithmic scale of the ordinate) from repeated collisions.
Notably, growth significantly accelerates as $\alpha$ increases, indicating that
higher local turbulence intensity (as represented here) effectively enhances
collision-coalescence.
The most pronounced changes occur when moving from $\alpha = 1.23$ to $\alpha = 1.63$ and $\alpha = 2.15$, which correspond to increases in the mean Stokes number from $\langle St \rangle = 0.23$ to $0.40$ and $0.69$, respectively.
This range marks the regime of strongest enhancement, consistent with the
observations from \cref{sec:monodisperse_suspensions}.

Gravity affects the DSD evolution primarily at smaller $\alpha$.
For $\alpha = 1$ and $\alpha = 1.23$, differential settling enhances collisions,
leading to faster broadening when gravity is included.
For larger $\alpha$, the DSDs with and without gravity become nearly indistinguishable.
This suggests that while differential settling is vital for typical cloud droplets under mean-field conditions, inertial effects become the dominant driver of collision-coalescence as the local dissipation rate increases.

\Cref{fig:dsdevolution}(b) shows the fraction of droplets that
undergo collision and coalescence as it evolves in time.
Collision activity ramps up during the first stage of the simulations, and
stabilizes after around $t=100\tau_\eta$.
Alongside the baseline simulations (solid lines), which are initialized with
polydisperse populations and include gravity, we also report reference results
from simulations without gravity (dashed) and from simulations initialized with
monodisperse populations (dotted; gravity included).
The suspensions are extremely dilute, and consequently only a minute fraction of
droplets, on the order of $10^{-6}$ to $10^{-5}$, undergo collision and coalescence.
Nevertheless, the collision rate increases sharply with $\alpha$.
Relative to the $\alpha=1.23$ case,
collisions occur about $3.8$ times more frequently with $\alpha=1.63$,
supporting the notion that intermittent events of intense turbulence can
substantially accelerate droplet growth.
The impact of gravity can be quantified more precisely here:
We observe that differential settling enhances collision-coalescence for
$\alpha=1.23$ by $34\%$, whereas its effect is negligible in the
$\alpha=1.63$ case (dashed and solid lines coincide in \cref{fig:dsdevolution}(b)).

Finally, the data highlight the complex role of polydispersity.
At lower dissipation levels ($\alpha=1.23$), polydispersity
enhances droplet growth substantially, yielding roughly $2.2$ times more
collisions compared to the corresponding monodisperse simulation.
This enhancement, however, weakens as the importance of droplet inertia
increases, reflecting our findings from~\cref{sec:bidisperse_suspensions}, and
because gravitational settling becomes relatively less important.
For $\alpha = 1.63$, collision-coalescence in the polydisperse case
exceeds that in the monodisperse reference by only about $3\%$.

\section{Conclusion} \label{sec:conclusion}
We have studied polydisperse droplet collisions in homogeneous isotropic
turbulence at Reynolds numbers up to $Re_\lambda=418$ by means of direct
numerical simulations (DNSs).
The bidisperse collision kernel, relative radial velocity (RRV) and radial
distribution function (RDF) at contact have been obtained for droplet pairs
with Stokes numbers in the range $St\in[0.02,2]$.
These data are important for understanding how turbulence affects the growth of
droplets in warm atmospheric clouds.
The statistics were obtained in the absence of gravity, in order to isolate
turbulent mechanisms and to target the smallest droplets within the bottleneck
range.
While these results are particularly relevant in the context of recent theories
of warm rain formation that invoke turbulent intermittency
and rapid growth of statistically ``lucky'' droplets,
it is important to bear in mind that they cannot be used directly in
large-eddy simulations (LESs) of convective clouds without accounting for
gravitational settling of larger droplets.

Our findings establish that differential sampling is an effective mechanism by
which differently-sized droplets collide under turbulence, even in the absence
of gravity.
This contrasts with the conventional view that primarily links polydispersity
to collisions induced by gravitational settling.
Droplets with separate relaxation times cluster less, but exhibit larger relative
velocities, so that the influence of polydispersity is highly dependent on the
Stokes number.
At $St \lesssim 0.4$, polydispersity enhances collisions because the increased
relative velocities from differential sampling compensate for the reduction in
clustering.
Conversely, at larger Stokes numbers, preferential concentration and the sling
effect are the dominant collision mechanisms, and polydispersity attenuates
collisions in this regime because it rapidly decorrelates the number density
fields.
Thus, the role of polydispersity transitions from enhancing to
surpressing collisions as droplet inertia increases.

We have compared our DNS data to the Onishi
model~\cite{onishiReynoldsnumberDependenceTurbulence2016} and found good
agreement for the monodisperse collision kernel up to $St\approx1.5$.
In contrast, the Zhou model~\cite{zhouModellingTurbulentCollision2001},
providing an expression
for the spatial correlation between differently-sized droplets,
was shown to introduce systematic errors in the bidisperse RDF, since it
overestimates the overlap of the droplet clusters as the Stokes numbers
separate.
This mismatch was observed for all simulated Reynolds numbers
($Re_\lambda=55$, $179$ and $418$),
and may originate from the use of frozen flow fields in the original study.
Furthermore, we found that the dispersion parameter $\theta$ (as defined in
\cref{eqn:theta}) is a better choice to model the correlation
coefficient, and we provided a modified expression (\cref{eqn:rho_theta}) that
fits the DNS data well across all Reynolds numbers.

We have introduced a new compact expression for the bidisperse collision kernel.
This parameterization originates from the observation that, although the
underlying bidisperse RRV and RDF display difficult-to-model variability with
respect to $St_i$ and $St_j$, the collision kernel in itself varies
smoothly with the Stokes number separation.
By exploiting symmetries and identifying the dispersion parameter $\theta$
as the relevant measure, our model results from a first-order approximation
that reproduces the DNS reference with high accuracy and remains
robust across Reynolds numbers.
Although differential settling strongly impacts
polydisperse collisions in the presence of gravity, we show that the proposed
parameterization remains qualitatively predictive, reproducing the collision
kernel at $Fr = 0.173$ within acceptable accuracy.
A more systematic exploration spanning a broader range of, and lower, Froude
numbers is deferred to future work.

Simulations of coalescing droplets have shown that cloud droplet size
distributions (DSDs) broaden within short time intervals, even for
typical cloud droplets with Stokes numbers around $St \approx 0.15$, providing a
viable pathway for growth within the bottleneck range.
A small fraction of lucky droplets grow disproportionately fast.
Whether this arises from mere statistical chance or specifically due to the
turbulent flow topology is the subject of future studies.
By emulating fluctuations of the local dissipation rate, we found that droplet
growth is markedly accelerated in highly dissipative parcels,
indicating that intermittency may play an important role in the warm rain
process.
Depending on droplet size and local turbulence intensity, initially narrow DSDs
can broaden as rapidly as wider ones, and at sufficiently elevated dissipation
rates, gravitational effects can become practically negligible, underscoring
the complex and nontrivial coupling between local flow conditions, the initial
droplet population, and its subsequent evolution.
Overall, these results support the long-standing hypothesis that turbulent
intermittency may contribute to overcoming the bottleneck problem; however, our
evidence is limited to Reynolds numbers smaller than those in actual clouds.
Numerical simulations are unlikely to reproduce turbulent flows that rival
atmospheric conditions in the foreseeable future.
Given these constraints, complementary approaches such as high-resolution
holographic imaging in clouds appear most promising to clarify this issue.

\begin{acknowledgments}
    This research was supported by funding from the Swiss National Science
    Foundation (grant 204621).
    Numerical simulations were carried out using the resources provided by the
    Euler cluster of ETH Z\"urich.
    We thank Arthur Couteau for providing the pseudo-spectral Navier-Stokes DNS solver
    used in this work.
\end{acknowledgments}

\section*{Author contributions statement}
L.A.C.
performed the simulations.
All authors analyzed the results.
D.W.M.
lead the acquisition of funding.
P.J.
supervised the project.
L.A.C.
wrote the original manuscript.
All authors reviewed the manuscript.

\appendix
\section{Onishi model for the monodisperse radial distribution function}
\label{sec:onishi_appendix}
The Onishi model~\cite{onishiReynoldsnumberDependenceTurbulence2016} for the
monodisperse radial distribution function is
\begin{equation}
    g_{ii}(R) - 1 =
    \begin{cases}
        A_1 St^2           & (\equiv y_1) \quad (\text{for } St_i < St_a)    \\ A_2
        Re_\lambda St^{-2} & (\equiv y_2) \quad (\text{for } St_a \leq St_i)
    \end{cases}
    ,
\end{equation}
with $A_1 = 110$, $A_2=0.38$ and
$St_a=((A_2/A_1)Re_\lambda)^{1/4}$.
% \label{eqn:onishi} 
The two formulations are connected in the composite expression
\begin{equation}
    g_{ii} - 1 = H(St - St_a) y_1 z_a^\alpha + H(St_a - St) y_2 (1 - z_a)^\alpha,
\end{equation}
where $H(\cdot)$ denotes the Heaviside function and
\begin{equation}
    z_a(St) = \frac{1}{2} \left( 1 - \tanh \frac{\log_{10} St -
        \log_{10} St_a}{C_a} \right)
\end{equation}
with $C_a=a_cRe_\lambda^{b_c}$,
$\alpha=\log_2(a_\alpha Re_\lambda^{b_\alpha})$, $a_c=0.046$, $b_c=0.36$,
$a_{c2}=0.094$, $b_{c2}=0.25$, $a_\alpha=0.23$ and $b_\alpha=0.5$.

\section{Wang model for the radial relative velocity}
\label{sec:wang_appendix}
The Wang model~\cite{wangStatisticalMechanicalDescription2000} for the radial
relative velocity is
\begin{equation}
    \langle w_r \rangle = \left[
        \frac{2}{\pi} \left( w^2_\text{shear} + w^2_\text{accel} \right) \right]^{1/2},
\end{equation}
with
\begin{subequations}
    \begin{align}
        w^2_\text{shear} & =
        \frac{R^2\eps}{15\nu}, \\ w^2_\text{accel} & = \frac{1}{3} C_w(\phi)
        f_\text{KK},           \\ f_{KK} & = \frac{\gamma u_{\text{rms}}^2}{\gamma - 1} \left\{
        (\xi_i + \xi_j) - \frac{4 \xi_i \xi_j}{(\xi_i + \xi_j)} \left[ \frac{1 + \xi_i
                + \xi_j}{(1 + \xi_i)(1 + \xi_j)} \right]^{1/2} \right\} \times \left[
            \frac{1}{(1 + \xi_i)(1 + \xi_j)} - \frac{1}{(1 + \gamma \xi_i)(1 + \gamma
                \xi_j)} \right],
    \end{align}
\end{subequations}
where $f_\text{KK}$ was adapted
from Kruis \& Kusters~\cite{kruisCollisionRateParticles1997},
$C_w(\phi)=1+0.6\exp\left[-(\phi-1)^{3/2}\right]$ was provided later on by Zhou
et al.~\cite{zhouModellingTurbulentCollision2001}, $\phi =
    \max(St_i/St_j,St_j/St_i)$, $\gamma = 0.183 u^2_\text{rms}/(\eps\nu)^{1/2}$ and
$\xi_i=\tau_{p,i}/(0.4u^2_\text{rms}/\eps)$.

\section{Collision map}
The map of collision kernels obtained from the DNS is provided in
\cref{tab:collision_map}.
\begin{table}[h!]
    \caption{Normalized bidisperse collision kernel $\Gamma_{ij}/\left[u_\eta R^2\right]$ for non-settling droplets mapped across Stokes number pairs.
        Data obtained from the DNS at $Re_\lambda=418$.}
    \label{tab:collision_map}
    \begin{ruledtabular}
        \begin{tabular}{l *{12}{S[table-format=1.5]}}
                   & \multicolumn{12}{c}{$St_i$}                                                                                                                                     \\
            \cmidrule(lr){2-13}
            $St_j$ &
            {0.02} & {0.04}                      & {0.06}    & {0.08}    & {0.10}    & {0.20}    & {0.30}    & {0.40}    & {0.50}    & {0.60}    & {0.70}    & {0.80}                \\
            \midrule
            0.02   & 0.012770                                                                                                                                                        \\
            0.04   & 0.165050                    & 0.031557                                                                                                                          \\
            0.06   & 0.276010                    & 0.180599  & 0.045403                                                                                                              \\
            0.08   & 0.428492                    & 0.326448  & 0.195588  & 0.057096                                                                                                  \\
            0.10   & 0.558481                    & 0.435732  & 0.382241  & 0.224400  & 0.082339                                                                                      \\
            0.20   & 1.185721                    & 1.153545  & 1.082863  & 1.022244  & 0.979117  & 0.379158                                                                          \\
            0.30   & 1.758476                    & 1.768431  & 1.779228  & 1.774839  & 1.760120  & 1.530779  & 1.122244                                                              \\
            0.40   & 2.332095                    & 2.335571  & 2.379300  & 2.354239  & 2.438115  & 2.493523  & 2.408844  & 2.324684                                                  \\
            0.50   & 2.803172                    & 2.904109  & 2.946676  & 2.985315  & 3.082578  & 3.311896  & 3.471911  & 3.574872  & 3.832678                                      \\
            0.60   & 3.348263                    & 3.379465  & 3.504257  & 3.620609  & 3.672195  & 4.035877  & 4.355600  & 4.671988  & 4.923528  & 5.470794                          \\
            0.70   & 3.777948                    & 3.899344  & 4.035190  & 4.128008  & 4.242323  & 4.687616  & 5.127529  & 5.535844  & 5.949110  & 6.327964  & 7.027269              \\
            0.80   & 4.231069                    & 4.290481  & 4.476302  & 4.635393  & 4.748258  & 5.284609  & 5.801385  & 6.277354  & 6.761310  & 7.240590  & 7.696198  & 8.476907  \\
            0.90   & 4.645948                    & 4.797752  & 4.958269  & 5.089479  & 5.231343  & 5.842295  & 6.393635  & 6.940931  & 7.458380  & 7.973970  & 8.478794  & 8.957733  \\
            1.00   & 5.043913                    & 5.234724  & 5.407762  & 5.491839  & 5.710500  & 6.378543  & 6.944720  & 7.508725  & 8.072345  & 8.606343  & 9.122944  & 9.626397  \\
            1.10   & 5.474131                    & 5.612824  & 5.787222  & 5.994847  & 6.140369  & 6.844883  & 7.475522  & 8.050023  & 8.606799  & 9.164909  & 9.683897  & 10.197659 \\
            1.20   & 5.870312                    & 6.071453  & 6.299756  & 6.417183  & 6.561811  & 7.300540  & 7.941470  & 8.551744  & 9.107339  & 9.655390  & 10.175764 & 10.681971 \\
            1.30   & 6.263289                    & 6.429219  & 6.650576  & 6.803644  & 6.967690  & 7.732056  & 8.386697  & 8.991832  & 9.557596  & 10.105922 & 10.622003 & 11.115861 \\
            1.40   & 6.569969                    & 6.787741  & 7.010799  & 7.251224  & 7.374559  & 8.137442  & 8.813926  & 9.422414  & 9.981835  & 10.527532 & 11.030228 & 11.533121 \\
            1.50   & 6.955757                    & 7.137685  & 7.352395  & 7.568374  & 7.732348  & 8.530612  & 9.210128  & 9.804870  & 10.364695 & 10.899514 & 11.416203 & 11.895059 \\
            1.60   & 7.244628                    & 7.512749  & 7.764366  & 7.935321  & 8.103869  & 8.896406  & 9.585476  & 10.185557 & 10.737608 & 11.270002 & 11.751157 & 12.243577 \\
            1.70   & 7.587673                    & 7.833124  & 8.114156  & 8.279690  & 8.436587  & 9.255811  & 9.936971  & 10.527685 & 11.088232 & 11.596512 & 12.088243 & 12.569693 \\
            1.80   & 7.956461                    & 8.152330  & 8.411341  & 8.595010  & 8.778074  & 9.589369  & 10.259292 & 10.859810 & 11.427766 & 11.920917 & 12.415937 & 12.861273 \\
            1.90   & 8.276954                    & 8.477343  & 8.717392  & 8.892874  & 9.103686  & 9.920902  & 10.592037 & 11.203534 & 11.730887 & 12.238757 & 12.716814 & 13.161478 \\
            2.00   & 8.579086                    & 8.807376  & 9.040300  & 9.243208  & 9.405027  & 10.230625 & 10.913011 & 11.507242 & 12.040707 & 12.545216 & 13.000854 & 13.450260 \\

            \toprule

                   & \multicolumn{12}{c}{$St_i$}                                                                                                                                     \\
            \cmidrule(lr){2-13}
            $St_j$ &
            {0.90} & {1.00}                      & {1.10}    & {1.20}    & {1.30}    & {1.40}    & {1.50}    & {1.60}    & {1.70}    & {1.80}    & {1.90}    & {2.00}                \\
            \midrule
            0.90   & 9.760407                                                                                                                                                        \\
            1.00   & 10.097957                   & 10.902536                                                                                                                         \\
            1.10   & 10.670798                   & 11.133210 & 11.892350                                                                                                             \\
            1.20   & 11.154569                   & 11.606633 & 12.047276 & 12.761468                                                                                                 \\
            1.30   & 11.586278                   & 12.034813 & 12.467982 & 12.865994 & 13.495080                                                                                     \\
            1.40   & 12.002002                   & 12.420590 & 12.840204 & 13.221509 & 13.591282 & 14.167918                                                                         \\
            1.50   & 12.350899                   & 12.776317 & 13.170369 & 13.540581 & 13.888741 & 14.235115 & 14.751990                                                             \\
            1.60   & 12.672636                   & 13.109496 & 13.502077 & 13.854756 & 14.202523 & 14.500906 & 14.812924 & 15.300914                                                 \\
            1.70   & 12.999646                   & 13.403924 & 13.786960 & 14.133588 & 14.474330 & 14.776241 & 15.055299 & 15.325259 & 15.756258                                     \\
            1.80   & 13.295512                   & 13.695877 & 14.060142 & 14.411927 & 14.732263 & 15.023991 & 15.307945 & 15.556875 & 15.829575 & 16.225749                         \\
            1.90   & 13.586718                   & 13.976292 & 14.325502 & 14.675783 & 14.984001 & 15.273409 & 15.540298 & 15.800630 & 16.030853 & 16.285003 & 16.626343             \\
            2.00   & 13.841518                   & 14.231365 & 14.585027 & 14.906587 & 15.222230 & 15.508153 & 15.769918 & 16.025453 & 16.242491 & 16.468293 & 16.695717 & 16.996604 \\
        \end{tabular}
    \end{ruledtabular}
\end{table}

\bibliography{ref}

\end{document}